\documentstyle[twoside,fleqn,espcrc2,epsf]{article}
\newcommand{\be}{\begin{equation}}
\newcommand{\ee}{\end{equation}}
\newcommand{\bea}{\begin{eqnarray}}
\newcommand{\eea}{\end{eqnarray}}
\newcommand{\bean}{\begin{eqnarray*}}
\newcommand{\eean}{\end{eqnarray*}}

\newcommand{\AmS}{{\protect\the\textfont2
  A\kern-.1667em\lower.5ex\hbox{M}\kern-.125emS}}

\title{Relativistic bottomonium spectrum from anisotropic lattices
\thanks{
This work was conducted on the QCDSP machines at 
Columbia University and RIKEN-BNL Research Center. 
TM and XL are supported by DOE. 
}}

\author{ 
X.~Liao\address{Physics Department, Columbia University, 
New York, NY 10027, USA} and 
T.~Manke$^{\rm a}$ }
\begin{document}

\begin{abstract}
We report on our results from a fully relativistic simulation
of the quenched bottomonium spectrum.
Using an anisotropic formulation of Lattice QCD,
we were able to retain a very fine resolution into the
temporal direction for a range of different lattice spacings.
At fixed renormalized anisotropy we study the scaling properties
of the spectrum and compare our results with non-relativistic
calculations.
\end{abstract}

%
\maketitle
\section{INTRODUCTION}
The non-perturbative study of heavy quark systems is complicated
by the large separation of momentum scales which are difficult to
accommodate on conventional isotropic lattices.
Several approximations to relativistic QCD have been proposed 
to describe accurately the low energetic phenomenology of 
heavy quarkonia \cite{nrqcd,fermilab}. 
However, the \mbox{(non-)}perturbative control of systematic
errors in those approximations is very difficult
and in practise one has to rely on additional approximations. 
The high precision results for the spin structure in 
non-relativistic bottomonium calculations have been hampered by large 
systematic errors which are difficult to control within this effective theory.
Higher order radiative and relativistic corrections are sizeable for 
bottomonium and even more so for charmonium.
Still more cumbersome are large observed scaling violations 
\cite{nrqcd_nf2} that cannot be controlled by taking the continuum limit.

We take this as our motivation to study heavy quarkonia on anisotropic 
lattices in a fully relativistic framework. This approach has previously 
been used to calculate the charmonium spectrum with unprecedented accuracy 
\cite{tim,ping,lat00} . 
Here we extent the method to even heavier quarks and focus on our new results
for bottomonium. Preliminary results were reported in \cite{lat00}.
In Section \ref{sec:actions} we briefly review our action and simulation 
parameters and Section \ref{sec:results} discusses our results.
\section{ANISOTROPIC LATTICES}
\label{sec:actions}
To study the relativistic propagation of heavy quarks it is mandatory
to have a fine resolution in the temporal lattice direction.
To this end we employ an anisotropic gluon action:

\bea
S = - \beta \left( \sum_{x, {\rm i > j}} \xi_0^{-1} P_{\rm i j}(x) +
    \sum_{x, {\rm i}}     \xi_0      P_{\rm i t}(x) \right) ~~.
\label{eq:aniso_glue}
\eea

This is the standard Wilson action written in terms of simple plaquettes,
$P_{\mu\nu}(x)$. Here $(\beta,\xi_0)$ are two bare parameters, which 
determine the gauge coupling and the renormalized anisotropy, 
$\xi = a_s/a_t$, of the quenched lattice. 
The anisotropic gluon action (\ref{eq:aniso_glue}) is designed to be accurate 
up to ${\cal O}(a_s^2,a_t^2)$. 

For the heavy quark propagation in the gluon background 
we used the ``anisotropic clover'' formulation as first described in 
\cite{tim,ping}. The discretized form of the continuum Dirac operator, 
$Q=m_q+D\hskip -0.21cm \slash $, reads
\bea
Q & = & m_0 + \nu_s~W_i \gamma_i + \nu_t~W_0 \gamma_0 - \nonumber \\ 
& & \frac{a_s}{2}\left[ c_s~\sigma_{0k}F_{0k} + c_t~\sigma_{kl}F_{kl} \right]~~, 
\nonumber \\
W_\mu & = & \nabla_\mu - (a_\mu/2) \gamma_\mu \Delta_\mu ~~.
\label{eq:aniso_quark}
\eea
The five parameters in Eq. \ref{eq:aniso_quark} are all related
to the quark mass, $m_q$, and the gauge coupling as they appear in the
continuum action. Their classical estimates have been given in \cite{ping}.
Here we chose $m_0$ non-perturbatively, such that the rest energy
of the hadron corresponds to its experimental value
($M(^{3}S_1^{--})$ = 9.46 GeV). We also fix $\nu_s=1$
and adjust $\nu_t$ non-perturbatively for the mesons to obey a relativistic dispersion relation (~$c({\bf 0})=1$~):
\bea
E^2({\bf p}) &=& E^2({\bf 0}) + c^2({\bf p})~{\bf p}^2 + {\cal O} ({\bf p}^4) \ldots ~~.
\eea

For the clover coefficients $(c_s,c_t)$ we choose their classical estimates
and augment this prescription by tadpole improvement. The tadpole coefficient
has been determined from the average link in Landau gauge: 
$u_{\mu} = 1/3~\langle U_\mu(x) \rangle_{\rm Landau}$. 
Any other choice will give the same continuum limit, 
but with this prescription we expect only small
${\cal O}(\alpha a)$ discretization errors. 

Using the anisotropic gluon action (\ref{eq:aniso_glue}) we generate several thousand
quenched gauge field configuration for a variety of lattice spacings and anisotropies.
Every few hundred sweeps we calculate relativistic quark propagators from the
inverse of Eq. \ref{eq:aniso_quark}. This leaves us with several hundred independent
measurements for each set of bare parameters.
The construction of meson correlators with definite $J^{PC}$ from quark propagators 
is standard \cite{lat00}.
Our fundamental bilinears are constructed from two 4-spinors, one of 
16 $\Gamma$- matrices and additional insertions of up to 2 lattice 
derivatives:  
\be
M(x) = \bar q(x) \Gamma_i \Delta_j \Delta_k q(x)~~.
\ee
We have further improved those operators to give states with
different projection onto the ground state. 
Here we follow \cite{nrqcd_hybrid} and employ a combination
of various quark smearings and APE-smearing for the gauge links. 
In practise we find it often useful to use gauge-fixed box sources,
where we can vary the overlap by simply varying the extent of the box.
This allows us to extract reliably both the ground state energies and 
their excitations from correlated multi-state fits to several channels.
We demand consistency for fit results from different fit methods and fit
ranges. In order to call a fit acceptable we require its Q-value to be bigger than 0.1.
In Table \ref{tab:para} we list the parameters of our simulation.
\begin{table}[b]
\vskip -7mm
\caption{Simulation Parameters. Using ${^1P}_1-{^3S}_1$ to set the scale, we adjust 
the bare quark mass such that the $1^{--}$ corresponds to the 
experimental values in $b \bar b (\Upsilon)$.}
\begin{tabular}{cccc}
\hline
$(\beta,\xi)$ & $N_s^3 \times N_t$ & configs    & $1^{--}$ [GeV]  \\
\hline
(5.7, 4)      & $8^3 \times 64$    & 291        & 10.58(67)       \\
(5.9, 4)      & $8^3 \times 64$    & 600        & 9.25(61)        \\
(6.1, 4)      & $16^3 \times 96$   & 660        & 10.40(30)       \\
(6.3, 4)      & $16^3 \times 128$  & 450        & 12.60(67)       \\
(6.5, 4)      & $16^3 \times 160$  & 710        & 13.04()         \\
(5.9, 5)      & $8^3 \times 96$    & 780        &  9.57(44)        \\
(6.1, 5)      & $16^3 \times 128$  & 370        & 10.11(97)        \\
\hline
\end{tabular}
\label{tab:para}
\end{table}

\section{BOTTOMONIUM SPECTRUM}
\label{sec:results}
\begin{figure}[t]
\hbox{\epsfxsize = 70mm  \epsfysize = 60mm \hskip -1mm \epsffile{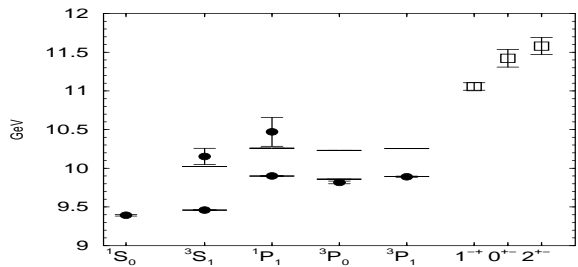}}
\caption{Quenched bottomonium spectrum. We plot continuum estimates as full circles. Selected results for exotic candidates (hybrids) from finite lattice spacing are shown as open squares. 
The ${^1P}_1-{^3S}_1$ splitting is used to set the scale.}
\label{fig:continuum}
\end{figure}

In Figure \ref{fig:continuum} we show the continuum limit of our 
results and selected exotic candidates in the bottomonium spectrum.

The later are results from only one lattice spacing at (6.1,4).
To convert our lattice results into dimensionful quantities we
used the ${^1P}_1-{^3S}_1$ splitting which is $\approx 440$ MeV experimentally.
It is well-known that, without dynamical sea quarks in the gluon background, 
the definition of the lattice spacing is ambiguous and one cannot 
reproduce all experimental splittings simultaneously.
For the purpose of this work we admit to the uncertainties of the
quenched calculation and focus on other systematic errors instead.

As mentioned above we are tuning $\nu_t$ non-perturbatively at each mass,
so as to reproduce a relativistic dispersion relation to within $1-2$\%.
We tested for systematic errors on spectral quantities as we vary $\nu_t$
and find that they are small, $<1\%$. 
We have also performed a finite volume study and find that volumes with
$L\approx 1$ fm are sufficiently large to accommodate the small bottomonium
ground states. For even smaller volumes we could observe some shifts in the spectrum,
namely the hyperfine splitting is suppressed.

\begin{figure}[t]
\hbox{\epsfxsize = 70mm  \epsfysize = 60mm \hskip -1mm \epsffile{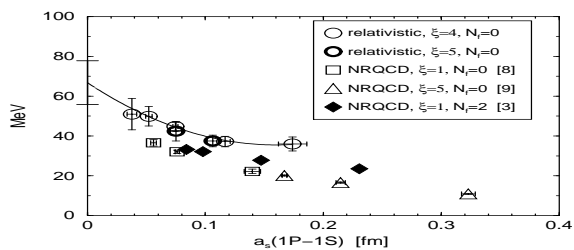}}
\caption{Hyperfine splitting. We show the continuum extrapolation
from a linear-plus-quadratic ansatz in $a_s$.}
\label{fig:hfs_vs_as}
\end{figure}

The most interesting aspect of our work is the determination of the spin-structure
in bottomonium from a relativistic formulation.
In Fig. \ref{fig:hfs_vs_as} we show the continuum extrapolation
of the hyperfine splitting along with selected results from 
non-relativistic calculations. 
Clearly there are large scaling violations which result
in a rather large continuum limit with sizeable errors.
We also notice some deviation from quenched non-relativistic calculations
at finite lattice spacing. We interpret those discrepancies as due to
higher order relativistic corrections as well as a different tadpole
prescriptions in \cite{nrqcd_scaling}. In Refs. \cite{nrqcd_nf2,mont99}
improved gluon actions were used which can also result in different
scaling violations.
From potential models and results in \cite{nrqcd_nf2}, we should expect 
dynamical calculations to raise the hyperfine splitting even further.
We also like to point out the very good agreement for our
simulations with different renormalized anisotropy (4 and 5). This indicates 
that we are working already close to the Hamiltonian limit of lattice QCD.

\begin{figure}[t]
\hbox{\epsfxsize = 70mm  \epsfysize = 60mm \hskip -1mm \epsffile{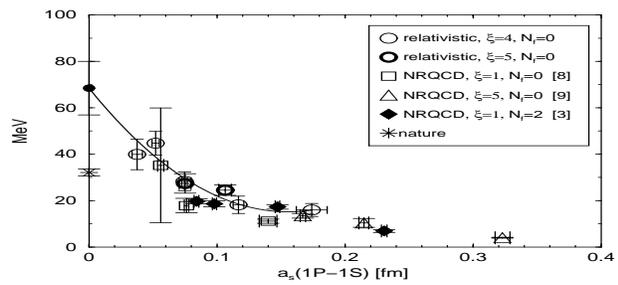}}
\caption{Fine structure. As in Fig. \ref{fig:hfs_vs_as}. }
\label{fig:fs_vs_as}
\end{figure}

For the fine structure, Fig. \ref{fig:fs_vs_as}, we observe a closer
agreement with NRQCD calculations. However, the large scaling violations
result in an overestimate of the experimental value by $3\sigma$.

For the first excited states, $2S$, we also find an enlargement
compared to experiment and in line with previous estimates of this 
quantity. We feel that anisotropic lattices provide a more reliable tool
to resolve such high excitation, but we leave a more detailed analysis
to future work. 

In conclusion, we have demonstrated that a relativistic 
simulation of bottom quarks is possible on anisotropic 
lattices. Work is in progress to treat also light dynamical
quarks in an anisotropic fashion.

\end{document}